\documentclass[a4paper,twocolumn]{article}

\usepackage{jsaiac}
\usepackage{color}
\usepackage[dvipdfmx]{graphicx}
\usepackage{multirow}

\title{A neuro-symbolic approach for\\ multimodal reference expression comprehension}

\address{Aman Jain, jain-aman@g.ecc.u-tokyo.ac.jp}

\author{%
Aman Jain\first\second
\and
Anirudh Reddy Kondapally\first\second
\and
Kentaro Yamada\second
\and
Hitomi Yanaka\first
}

\affiliate{
\first{} The University of Tokyo
\and
\second{} Honda R\&D Co.,Ltd., Tokyo, Japan
}

\begin{abstract}
Human-Machine Interaction (HMI) systems have gained huge interest in recent years, with reference expression comprehension being one of the main challenges. Traditionally human-machine interaction has been mostly limited to speech and visual modalities. However, to allow for more freedom in interaction, recent works have proposed the integration of additional modalities, such as gestures in HMI systems. We consider such an HMI system with pointing gestures and construct a table-top object picking scenario inside a simulated virtual reality (VR) environment to collect data. Previous works for such a task have used deep neural networks to classify the referred object, which lacks transparency. In this work, we propose an interpretable and compositional model, crucial to building robust HMI systems for real-world application, based on a neuro-symbolic approach to tackle this task. Finally we also show the generalizability of our model on unseen environments and report the results.
\end{abstract}

\def\BibTeX{{\rm B\kern-.05em{\sc i\kern-.025em b}\kern-.08em%
 T\kern-.1667em\lower.7ex\hbox{E}\kern-.125emX}}
\def\JBibTeX{\leavevmode\lower .6ex\hbox{J}\kern-0.15em\BibTeX}
\def\LaTeXe{\LaTeX\kern.15em2$_{\textstyle\varepsilon}$}

\begin{document}
\maketitle

\section{Introduction}
\label{sec:intro}
With the breakthrough successes in visual recognition and linguistic understanding, multimodal inference tasks involving both linguistic and visual modalities are being challenged. Reference expression comprehension (REC) task\cite{liu2019clevr,yu2016modeling} is one such task which involves identifying an object being referred in the image from a natural language query. It is a particularly challenging problem since it requires understanding the language semantics and matching them with visual information like color, shape or name of objects. The REC task has wider practical applications as well. One of them being in human-machine interaction (HMI) systems capable of sensing multiple modalities --- speech and vision.

\begin{figure*}[h!]
\centering
\includegraphics[width=\linewidth]{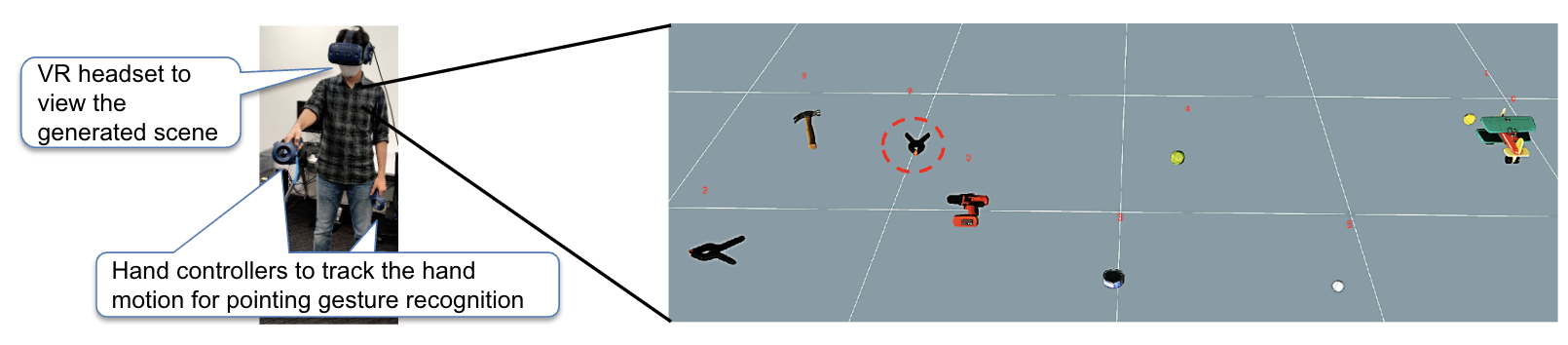}
\caption{Data collection setup. The image on left is the HTC Vive VR setup with head and hand controllers. The image on right is the simulated environment visible from inside the VR headset. In this example, the annotator records the utterance --- ``Pick up the black clipper beside this tool" to refer to the object marked with the dotted red circle.} 
\label{fig:dataset_collection}
\end{figure*}

In this work, we explore the REC task in the context of such a multi-modal HMI system. Specifically, we consider a robot which shares an environment with its user. The robot needs to carefully understand its surrounding and the objects inside it. The user then gives an instruction to the robot involving an object. The robot needs to identify the object being referred to by its user in order to fetch it. However, in daily life, humans also tend to use other nuances like gestures and gaze to provide additional cues in the instructions, as can be seen in Figure \ref{fig:dataset_collection} where the user uses pointing gesture to refer to the drill machine as an intermediate step while describing the clipper. In this work, we include one such modality, the pointing gesture, so the user can provide the instruction using a combination of pointing gesture and linguistic utterance simultaneously.

The current state-of-the-art methods for multimodal inference tasks include several deep neural networks, such as joint transformer-based architectures\cite{lu2019vilbert, chen2020uniter}. They have proven to be successful in various downstream multimodal inference tasks. These architectures work well in grounding the attributes and objects but lack the power to perform complex reasoning and are  opaque in their reasoning. On the other hand, symbolic program execution and reasoning have proven to be very powerful in performing complex reasoning but lack the scalability and generalizability of the former methods. To bridge this issue, some recent works have introduced a hybrid approach of marrying both these methods\cite{hudson2019learning, yi2018neural, mao2018the}. These methods, also known as neuro-symbolic methods, first use the deep neural networks to parse the raw visual and linguistic features into symbolic structures. Then, these symbolic structures are fed to a reasoning and execution module to arrive at the solution, enabling to develop an interpretable and compositional model. 
%This neuro-symbolic approach also enables to develop an interpretable and compositional model which is crucial for building robust HMI systems that can be deployed in the real-world.

In this work, we propose a neuro-symbolic approach to the REC task in an HMI system involving linguistic and pointing gesture based instructions. We construct a toy dataset for an object picking HMI system in a VR simulated environment to evaluate our method. Finally, we test our model's ability to generalize to unseen environment as well.

\section{Related Works}
\label{sec:related}
Referring expression comprehension on images is a very well researched area with several datasets\cite{yu2016modeling,liu2019clevr}. The popular datasets for this task like RefCOCO and RefCOCO+\cite{yu2016modeling} mainly focus on grounding the objects and attributes in the linguistic query to the image regions. These works do not consider any gesture modality and often do not require any complex multi-hop reasoning. These limitations are overcome by a recent work, CAESAR\cite{islam2022caesar}, which consists of multi-modal reference expressions with pointing gestures in a simulator environment. However, there are a few key differences with our work. Firstly, their instructions, environment and pointing gestures are all generated automatically from templates, neglecting the natural variations in human utterance. Secondly, their architecture does not involve a separate reasoning module and hence is not a transparent and compositional module as ours, making it difficult to build an interpretable and robust HMI system crucial for practical applications. Another such work on reference expressions in HMI systems is M2Gestic\cite{10.1145/3382507.3418863}. Unfortunately, it does not truly utilize the pointing gesture in conjunction with the linguistic instruction while inferencing, but rather only at the end to pick the best from the potential candidates inferred using the other two modalities (vision and text) alone. Hence, it will struggle to perform multi-hop inference involving complex instructions of intertwining gestures and text instructions.

In terms of the model architecture, the most related work, which we have taken inspirations from, is the Neural State Machine (NSM)\cite{hudson2019learning}. It has proven to be very effective for the visual question answering (VQA)\cite{antol2015vqa} task, which is another multimodal inference task with the objective being to answer a question by looking at an image. We borrowed the core ideas of disentengled representations and state-machine based reasoning to develop an architecture for the REC task. Moreover, we included another modality of pointing gesture into this architecture, and incorporated various heuristics to reduce parameters so that it can be trained for a task with fewer training samples. 

\begin{figure*}[h!]
\centering
\includegraphics[width=\linewidth]{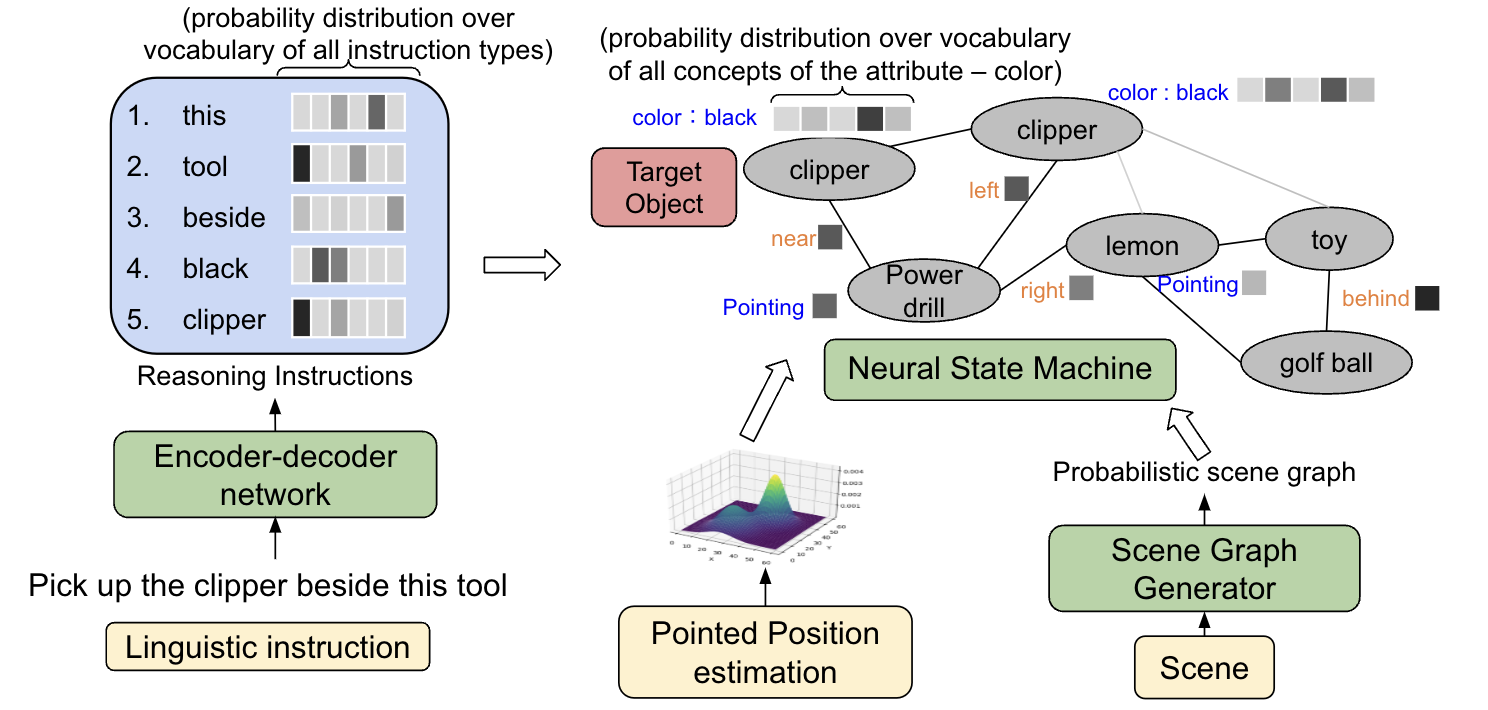}
\caption{Model architecture overview.}
\label{fig:model}
\end{figure*}

\section{Dataset}
\label{sec:dataset}

We constructed a virtual reality (VR) simulated environment to collect data for this task. The setup included an HTC Vive VR headset with hand controllers. For the simulated environment, we considered a scene with various objects placed on the floor. We create this environment in Rviz\cite{10.1007/s11235-015-0034-5} and use 3D object meshes from \cite{7251504} consisting of common everyday objects to sample objects. The task is then designed as follows. First, the user wears the HTC Vive headset and hand controllers, and looks in the simulated environment with objects placed on the ground. Then the user speaks the instruction to pick up one of the objects on the ground, also using the hand controllers to perform pointing gesture if required. %The machine needs to identify the object user asked it to pick up. 

% To populate the objects inside the simulated environment, we use the 3D object meshes from (TODO : cite source).
% We randomly choose 10 objects (with replacement) from this set and assign them a random position. We make sure that there is a minimum amount of distance between each pair of objects to ensure there is no overlapping in the objects and also each object is visible clearly. We use Rviz\cite{10.1007/s11235-015-0034-5}, a graphical interface for ROS, to generate the environment. 

%We then ask human annotators to wear the VR setup, look at the random generated scene and specify the instructions to pick up an object in the scene. We
For data collection we ask human annotators, the co-authors, to interact with our system and collect their linguistic instruction, the hand controllers' positions throughout the experiment and the ground truth object being referred to for each run. For example, as shown in Figure \ref{fig:dataset_collection}, a human annotator sees the simulated scene on the right from the VR headset and utters the instruction --- ``Pick up the black clipper beside this tool'' to refer to the clipper (highlighted with the red dashed circle) while performing pointing gesture using the hand controllers to point at the power drill.

We collect 130 samples which we split into 104 for training and rest 26 for validation. For the test set, we again collect 40 such samples but this time using a completely different set of objects from \cite{nouri:hal-01441721} consisting of various toys.
Using completely different objects for the test set helps in ensuring that our model actually generalizes to new environments and perform compositional reasoning required for understanding the reference expressions, not just relying on some statistical correlations.

\section{Method}
\label{sec:method}
Figure \ref{fig:model} shows an overview of our model architecture. The general idea is to transform each modality from a raw feature space to an abstract semantic space, then combine them inside a state machine to perform reasoning about the referred object. Each component of this model is explained in detail further in this section. \\

\subsection{Reasoning steps generator}
First, the input linguistic instruction is processed into a sequence of unit semantic instructions. One common way of doing this is by treating this task as a machine translation task and employing a deep neural encoder decoder network to generate the unit semantic instruction steps. However, this requires a huge amount of data to be trained, which is not the case with our dataset.

\begin{figure}[h!]
\centering
\includegraphics[width=\linewidth]{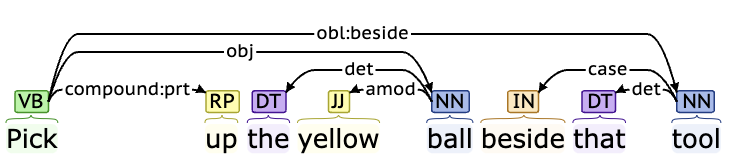}
\caption{Dependency parser output for a sample instruction.}
\label{fig:dep_parsing}
\end{figure}

So, we parse the input linguistic instruction by understanding the underlying grammatical patterns in the form of dependency parsing. Dependency parsers have been trained on vast amount of natural language data to extract the dependency relations between the semantic tokens in the sentence, and output a tree of dependencies as shown in Figure \ref{fig:dep_parsing}. We use the dependency parser from the Stanford CoreNLP toolkit\cite{manning-EtAl:2014:P14-5} for our purposes. We traverse this dependency tree in a depth-first order, heuristically choosing the sub-tree at each recursive step from the dependency relation type. Moreover, we filter the unnecessary tokens using stop-words and part-of-speech (POS) tag-based filtering. This traversal then generates the desired step-wise reasoning instructions to process the linguistic instruction.

We further classify each reasoning instruction step into one of the reasoning type categories : [object name, color, shape, size, demonstrative, relation]. These types are necessary to perform the step-by-step reasoning inside the state machine, which will be explained later in section \ref{subsec:NSM}. This classification is done heuristically by using the POS and dependency tags. However, if they do not fall into one of the patterns, the cosine similarity score between the instruction step and all the category embeddings is used for classification.

\subsection{Pointing Position estimation}
The wrist position and orientation in 3D space can be extracted from the HTC Vive's hand controller output. In each experiment, we collect the positions for both hands from beginning till the end. From these positions, we first extract the duration for which a pointing gesture is performed, if any, and then use the data in that duration to estimate the target pointed position on the ground. To identify a pointing gesture we use a simple heuristic --- if one of the hands is raised up in the air and the wrist trajectory (controller's trajectory) correspond to a straight arm motion, then that hand is in a pointing gesture motion.

Once we identify the gesture, the positions data corresponding to the pointing arm for the pointing duration is extracted. For each of these wrist positions, a gesture ray is constructed by joining it with the head controller's position. The intersection points of all these gesture rays on the ground plane is calculated and stored. These points on the ground are then used to construct a density estimation plot using Kernel density estimation (KDE) algorithm\cite{10.1214/aoms/1177728190,10.1214/aoms/1177704472}. The point with maximum density is then estimated as the target pointed position on the ground. This is based on the assumption that the user spends a large amount of time pointing directly at the target and lesser amount of time searching for the target or in motion. So the density of points should be highest around the target pointed position. Using this target pointed position and the density plot, each object is allocated a probability score of being the target pointed object. This score is then treated as one of the attributes of that object in the scene graph constructed as explained in the next subsection.

\subsection{Scene Graph Generator}
The scene graph generator module needs to transform the raw image pixels as perceived by the machine into a structured scene graph. A scene graph consists of nodes representing the objects in the scene and the edges representing the relations between those objects. The nodes also contain the following attribute information ---- name, color, shape and size. Due to the limited amount of data available for our task, it is not possible to train a reliable visual recognition model for scene graph generation. Hence, for the purpose of this work, we use an oracle scene graph generator using the ground truth object detections and their annotated attributes. The relations are classified amongst one of the spatial relations (left, right, front, back and near) using a simple heuristic on the relative position on the ground between each pair of object in the scene. Moreover, each attribute and relations in this scene graph is represented as a probability distribution over the vocabulary of each attribute and relation classes as done in \cite{hudson2019learning}
resulting in what is known as a probabilistic scene graph.

\subsection{Neural State Machine for reasoning} \label{subsec:NSM}
Once all the input modalities are transformed into their respective semantic structural space, we perform reasoning using a neural state machine as described in \cite{hudson2019learning}.
The probabilistic scene graph created from the scene is treated as a state machine on which the reasoning instruction steps are fed one-by-one to perform sequential reasoning. At each reasoning step, the state machine redistributes its probability distribution for the target node. Once all the steps are processed, the probability distribution on the state machine corresponds to that of the target object referred to by the input instruction.

The redistribution of probabilities at each step, $i$, is done by using the reasoning step $r_i$, its type $R_i$ (a tensor of size $L+1$ where $L$ is the number of attributes = 5), the nodes' representations $s^j$ for each attribute $j$ from 1 to $L$, and edges representations $e'$. Using these we first calculate a relevance score for each node ($\gamma_i(s)$) and each edge $\gamma_i(e)$ as follows :- %% TODO : update equation
\begin{equation}
\gamma_i(s) = \sigma ( \sum_{j=0}^L R_i(j)(r_i \circ W_j s^j) )
\end{equation}
\begin{equation}
\gamma_i(e) = \sigma ( r_i \circ W_{L+1} e' )
\end{equation}

Having these relevance scores, we compute the next states' probabilities of each node and edge independently.  
\begin{equation}
p_{i+1}^s = softmax_{s \in S}(p_i(s) \cdot \gamma_i(s))
\end{equation}
\begin{equation}
p_{i+1}^r = softmax_{s \in S} (\sum_{(s',s) \in E} p_i(s') \cdot \gamma_i((s',s)))
\end{equation}

Finally we combine them by a weighted sum, the weight being the probability of this reasoning step being of type relation, which is $R_i(L+1) = r_i'$
\begin{equation}
p_{i+1} = r_i' \cdot p_i^r + (1 - r_i') \cdot p_i^s
\end{equation}

Once all the reasoning steps are processed by the state machine, we get the final probability distribution of the target referenced object over all nodes in the graph. We use cross-entropy loss with the gold label node for training the model, and output the object corresponding to the node with highest probability score while inferencing. 

\section{Experiments}
\label{sec:experiment}
\begin{table}[]
\caption{Accuracy on validation and test set for different experiment settings}
\label{table:results}
\begin{tabular}{lcc}
\hline
\multicolumn{1}{c}{\multirow{2}{*}{Experiment Setting}} & \multicolumn{2}{c}{Accuracy (\%)} \\ \cline{2-3} 
\multicolumn{1}{c}{}                                    & Validation     & Test     \\ \hline
Normal (all input modalities)                           &                 80.8   &        72.5\\
Without pointing gesture input                          &                 23.1   &         10.0     \\ \hline
\end{tabular}
\end{table}

We train our model on the complete training set of 104 samples and evaluate on the validation set of 26 samples. Furthermore, we test the generalization ability of our model by evaluating it on the generalization test dataset created as explained in section \ref{sec:dataset}. We also perform ablation study in the input modalities to understand the importance of pointing gesture in our task. The results for all these experiments are summarized in Table \ref{table:results}. 

The comparable performance in validation and test set proves the generalization capabilities of our model, even with very low amount of training data. This is made possible because of the extensive compositionality and modularity in our method. We break down the input modalities into structured semantic concepts and perform reasoning on this abstract space, which makes it easier to train on relatively smaller amount of data with less trainable parameters and still be able to generalize to new environment. This is in contrast to the deep-learning approaches, which often struggle to generalize even with vast amount of training data and end up learning statistical biases on raw low-level feature space. 

From the ablation study results in the second row of the Table \ref{table:results}, we can conclude that pointing gesture is indeed very crucial for this task. The accuracy is significantly worse for the model which does not use the pointing gesture modality input. This is expected since humans often tend to convey implicit information through these gestures, which are very important to completely understand the instruction. 

\section{Conclusion}
\label{sec:conclusion}
In this work, we tackled the issue of reference expression comprehension for human-machine interaction systems, involving multiple modalities --- language, vision, and gestures. To do so, we collected a small but challenging dataset in a simulated VR environment to mimic a real-world application of such a task. Moreover, we proposed a novel method based on a neuro-symbolic approach to solving the REC task and showed its effectiveness in performing complex multi-hop reasoning. We also prove that such a method can generalize to unseen environments as well. Finally, we also performed ablation studies to emphasize on the importance of pointing gestures in real-world interactions tasks. 

However, much still remains to be done to build robust HMI systems capable of doing this task in real-world. First and foremost, a larger dataset on real-world HMI task focusing on reference expressions is required. Collecting such a data is a tedious task with monumental efforts but will reap rich rewards for the research in this field. Secondly, more sophisticated reasoning modules need to be developed to interpret various variety of reasoning such as ordinal references, counting, perspective disambiguation etc.

\subsection*{Acknowledgements}
This work was supported by JST, PRESTO Grant Number JPMJPR21C8, Japan.

\bibliographystyle{jsai}
\bibliography{ref}

\appendix
\end{document}